\def\rfr#1{eq. (\ref{#1})}

\def\bm#1{{\mbox{\boldmath$#1$\unboldmath}}}

\def\ctz#1{Ref.~\refcite{#1}}

\def\bm#1{{\mbox{\boldmath$#1$\unboldmath}}}
\def\bar{\begin{eqnarray}}
\def\ear{\end{eqnarray}}
\def\bb{\bibitem}
\def\eqI{\begin{equation}}
\def\eqF{\end{equation}}
\def\eqIa{\begin{eqnarray}}
\def\eqFa{\end{eqnarray}}
\def\rp#1#2{{#1\over#2}}

\def\lb#1{\label{#1}}

\def\oc2{$\mathcal{O}(c^{-2})$}

\documentclass{ws-ijmpd}

\begin{document}

\title{WILL IT BE POSSIBLE TO MEASURE INTRINSIC GRAVITOMAGNETISM  WITH LUNAR LASER RANGING?}

\author{LORENZO IORIO}

\address{INFN-Sezione di Pisa. Viale Unit$\grave{a}$ di Italia 68, 70125 Bari (BA), Italy
\\e-mail: lorenzo.iorio@libero.it}

 \maketitle

\begin{history}
\received{16 June 2008}
%\revised{Day Month Year}
\accepted{21 September 2008}
\comby{Jorge Pullin}
\end{history}

\begin{abstract}
In this note we mainly explore the possibility of measuring  the action of the intrinsic gravitomagnetic field of the rotating Earth on the orbital motion of the Moon with the Lunar Laser Ranging (LLR) technique. Expected improvements in it should push the precision in measuring the Earth-Moon range to the mm level; the present-day Root-Mean-Square (RMS) accuracy in reconstructing the radial component of the lunar orbit is about 2 cm; its harmonic terms can be determined at the mm level. The current uncertainty in measuring the lunar precession rates is about $10^{-1}$ milliarcseconds per year. The Lense-Thirring secular, i.e. averaged over one orbital period, precessions of the node and the perigee of the Moon induced by the Earth's spin angular momentum amount to $10^{-3}$ milliarcseconds per year yielding transverse and normal shifts of $10^{-1}-10^{-2}$ cm yr$^{-1}$. In the radial direction there is only a short-period, i.e. non-averaged over one orbital revolution, oscillation with an amplitude of $10^{-5}$ m.
Major limitations come also from some systematic errors induced by orbital perturbations of classical origin like, e.g., the secular precessions induced by the Sun and the oblateness of the Moon whose mismodelled parts are several times larger than the Lense-Thirring signal. The present analysis holds also for the Lue-Starkman perigee precession due to the multidimensional braneworld model by Dvali, Gabadadze and Porrati (DGP); indeed, it amounts to about $5\times 10^{-3}$ milliarcseconds per year.
\end{abstract}

%PACS\ 04.80.-y; 96.20.-n

\section{Introduction}
In the framework of the linearized weak-field and slow-motion
approximation of general relativity, the gravitomagnetic effects\cite{Rug,Scia04} are induced by the off-diagonal components $g_{0i},\
i=1,2,3$ of the space-time metric tensor\cite{Mash01,MashNOVA} which are
proportional to the components of the matter current density of the
source $j_i=\rho v_i$.

There are essentially two types of mass currents in gravity\cite{Kop06}.
The first type is induced by the rotation of the matter source
around its center of mass and generates the intrinsic
gravitomagnetic field which is closely related to the proper
angular momentum $\bm S$ (i.e. spin) of the rotating body. The other type is
due to the translational motion of the source and is responsible
for the extrinsic gravitomagnetic field.

A debate has recently arisen concerning the possibility of measuring some extrinsic gravitomagnetic orbital effects affecting the motion of the Earth-Moon system in the Sun's field with the Lunar Laser Ranging (LLR) technique\cite{Mur07a,Kop07,Mur07b,Sof08,Ciu08,Kop08}.
Another test of extrinsic gravitomagnetism concerning the deflection of electromagnetic waves by Jupiter in its orbital motion has been performed in a dedicated radio-interferometric experiment\cite{For08}.
%the interpretation of certain aspects of such a test raised a controversy\footnote{See on the WEB http://physics.wustl.edu/cmw/SpeedofGravity.html and %references therein.}.

In this brief note we wish to consider in some details the possibility of measuring with LLR an effect induced by the intrinsic gravitomagnetic field of the spinning Earth
\eqI \bm
B_g=\rp{G\left[3\bm r\left(\bm r\cdot \bm S \right)-r^2 \bm
S\right]}{cr^5}\lb{gmfield}\eqF
through the non-central, Lorentz-like acceleration \eqI\bm a = -2\left(\rp{\bm v}{c}\right)\times \bm B_g\lb{giusto}\eqF
on the orbital motion of the Moon around the Earth. In \rfr{gmfield} and \rfr{giusto} $G$ is the Newtonian gravitational constant, $c$ is the speed of light in vacuum, $\bm S$ is the Earth's spin angular momentum and $\bm v$ is the velocity of the Moon with respect to the Earth. The orbital feature we are interested in consists of the Lense-Thirring precessions of the longitude of the ascending node $\Omega$ and the argument of pericentre $\omega$ of the orbit of a test particle\cite{LT}. Twenty years ago, Bertotti in \ctz{Berto} wrote that the Lense-Thirring effect for the Moon was, at that time, too small to be detected; according to Ciufolini\cite{Ciu08}, intrinsic gravitomagnetism is still unmeasurable with the lunar orbit.  Instead, the possibility of measuring it in view of the expected forthcoming improvements in LLR  has recently been envisaged by M\"{u}ller {\it et al.} in \ctz{Mul07} 
and in \ctz{Mul06}; Kopeikin {\it et al.} in \ctz{KopASR} retain highly plausible its measurement.  %who write
%``With an improved accuracy the investigation of further effects (e.g. the Lense-Thirring precession) [...] become possible.''; according to %M\"{u}ller {\it et al.} in \ctz{Mul06}, with ``an improved accuracy of the LLR measurements and the modeling [...] the investigation of further %effects (e.g. the Lense-Thirring precession) [...] might become possible.''.
An overview of other attempts to measure the Lense-Thirring effect in various Solar System scenarios with natural and artificial test particles can be found in \ctz{IorNOVA}. In particular, for the LAGEOS-LAGEOS II test in the gravitational field of the Earth see \ctz{Ciu04} and \ctz{Ries}; for the perspectives on measuring the solar intrinsic gravitomagentic field with the inner planets of the Solar System see \ctz{LTVenus};  for the test performed with the Mars Global Surveyor in the field of Mars see \ctz{MGS} and \ctz{Krogh}.
Another effect induced by the intrinsic gravitomagentic field of the Earth is the precession of orbiting gyroscopes\cite{Pugh,Sch60} currently under measurement by the GP-B mission; see on the WEB http://einstein.stanford.edu/, \ctz{Eve} and \ctz{GPB}.

It is interesting to note that our analysis is equally valid also for the anomalous Lue-Starkman\cite{LS} perigee precession predicted in the framework of the multi-dimensional braneworld model of modified gravity put forth by Dvali, Gabadadze and Porrati\cite{DGP} (DGP) to explain the observed acceleration of the Universe without resorting to dark energy; indeed, as we will see, the magnitude of such an effect is the same as the Lense-Thirring one for the Moon. Several researchers\cite{LS,Zal,Tur08}  argued that it might be possible to measure the Lue-Starkman precession with LLR in view of the expected improvements in such a technique.

The physical and geocentric orbital parameters of the Moon are listed in In Table \ref{tavolaMoon}.
%
%------------------------------------------
\begin{table}
\tbl{Physical and geocentric orbital parameters of the Moon\protect\cite{Wil03,Ron05}. The gravity field adopted is the LP150Q solution (See on the WEB http://pds-geosciences.wustl.edu/geo/lp-l-rss-5-gravity-v1/lp$\_$1001/sha/jgl150ql.lbl).}
{\begin{tabular} {@{}cccc@{}}
\toprule
Parameter & Value &
Units \\
\colrule
\textit{m} mass& $7.349\times 10^{22}$ & kg \\
\textit{S} proper angular momentum& $2.32\times 10^{29}$ & kg m$^{2} s^{-1}$\\
\textit{Gm} & 4.902801076$\times 10^{12}$ & m$^{3}$ s$^{-2}$ \\
\textit{R} radius& $1.738\times 10^{6}$ & m\\
$\alpha$ proper angular velocity& 2.66$\times 10^{-6}$ & rad
s$^{-1}$\\
$\rp{C}{mR^2}$ normalized moment of inertia& $0.3932$ & -\\
{\it $J_2$} mass quadrupole moment & $2.0326\times 10^{-4}$ & -\\
{\it $\delta J_2$} uncertainty in the mass quadrupole moment & $1\times 10^{-8}$ & -\\
$a$ semi-major axis & $3.84400\times 10^8$ & m\\
$I$ mean inclination to the Earth's equator & $23.5$ & deg\\
$e$ eccentricity & 0.0549& - \\
\botrule
\end{tabular}\lb{tavolaMoon}}
\end{table}
%----------------------------------------------------------------

\section{The Lense-Thirring effect on the lunar orbit}
By assuming a suitably constructed geocentric equatorial frame, it turns out that  the node and the perigee of the Moon undergo the Lense-Thirring secular precessions
\begin{equation}\lb{blu}
\left\{
\begin{array}{lll}
\dot\Omega_{\rm LT}&=&\rp{2GS_{\oplus}}{c^2 a^3 (1-e^2)^{3/2}}=0.001\ {\rm mas\ yr}^{-1},\\\\
\dot\omega_{\rm LT}&=&-\rp{6GS_{\oplus}\cos I}{c^2 a^3 (1-e^2)^{3/2}}=-0.003\ {\rm mas\ yr}^{-1},
\end{array}
\right.
\end{equation}
where $a,e,I$ are the semi-major axis, the eccentricity and the inclination to the Earth's equator of the Moon's orbit;
mas yr$^{-1}$ stands for milliarcseconds per year. We used $S_{\oplus}=5.85\times 10^{33}$ kg m$^2$ s$^{-1}$\cite{IERS}.
The Lue-Starkman\cite{LS} pericentre precession is just about\footnote{The minus sign is related to the standard Friedmann-Lema\^{\i}tre-Robertson-Walker (FLRW) branch, while the plus sign is related to the self-accelerated branch which should be able to explain the observed acceleration of the Universe without resorting to dark energy\protect\cite{Zal}.} $\mp 0.005$ mas yr$^{-1}$.

Since the ratio of the mass of the Moon to that of the Earth
is\cite{Sta98}  $\mu = 0.0123000383$, one may argue that \rfr{blu}, which has been derived for a test-particle like, e.g., an artificial satellite, does not apply to the Earth-Moon system.
The intrinsic gravitomagnetic spin-orbit effects in the case of a two-body system with arbitrary masses $m_{\rm A}$ and $m_{\rm B}$ and spins $S_{\rm A}$ and $S_{\rm B}$ have been derived by Barker and O'Connell in \ctz{Bar75}, Damour in \ctz{Dam88}, Wex in \ctz{Wex}; for the sake of simplicity, we will reason in terms of the node.
In this case, the total node precession $\dot\Omega_{\rm tot}$ accounts for the spin-orbit contributions of both bodies and also for a spin-spin term.
The expression of the node precession of a body A is\cite{Bar75,Dam88,Wex}
\eqI \dot\Omega_{\rm A}=\left(\rp{3+x_{\rm A}}{2c^2}\right)\rp{G(m_{\rm A}+m_{\rm B})}{a^3(1-e^2)^{3/2}}\rp{S_{\rm A}}{m_{\rm A}},\ x_{\rm A}=\rp{m_{\rm A}}{m_{\rm A}+m_{\rm B}},\eqF
so that
\eqI\dot\Omega_{\rm tot} = \dot\Omega_{\rm A} + \dot\Omega_{\rm B}.\eqF
Let us pose
\eqI m_{\rm A}=m_{\oplus}\equiv M,\ m_{\rm B}=m_{\rm Moon}\equiv m;\eqF
thus,
it is possible to obtain
\eqI\dot\Omega_{\oplus} = \left(1+\rp{3}{4}\mu\right)\rp{2GS_{\oplus}}{c^2 a^3(1-e^2)^{3/2}},\lb{spinTERRA}\eqF
\eqI\dot\Omega_{\rm Moon} = \left(1+\rp{3}{4\mu}\right)\rp{2GS_{\rm Moon}}{c^2 a^3(1-e^2)^{3/2}};\eqF
recall that $\mu$ is the Moon/Earth mass ratio.
It results that the precession of \rfr{spinTERRA} is larger than the Lense-Thirring one of \rfr{blu} by the multiplicative factor $\left(1+\rp{3}{4}\mu\right)=1.0092$ yielding an error of $10^{-5}$ mas yr$^{-1}$, which is completely negligible (see Section \ref{errori}).
Concerning the precession due to the lunar spin, we have
\eqI \rp{\dot\Omega_{\rm Moon}}{\dot\Omega_{\oplus}}=\left(\rp{3+4\mu}{4+3\mu}\right)\rp{1}{\mu}\rp{S_{\rm Moon}}{S_{\oplus}}=2\times 10^{-3},\eqF
i.e. it is of the order of $2\times 10^{-6}$ mas yr$^{-1}$, which is negligible as well.
The amplitude of the spin-spin term is  proportional  to\cite{Bar75,Dam88,Wex}
\eqI \dot\Omega_{\rm SS}\propto -\rp{3}{2c^2}\sqrt{\rp{GM(1+\mu)}{a^7}}\rp{1}{(1-e^2)^2}\rp{S_{\oplus}}{M}\rp{S_{\rm Moon}}{m}=6\times 10^{-9}\ {\rm mas}\ {\rm yr}^{-1}.\eqF
Thus, we can conclude that the Lense-Thirring approximation is fully adequate for the Earth-Moon system.
\section{Some sources of error}\lb{errori}
Let us now examine some sources os systematic errors.
In regard to the potentially corrupting action of the mismodelling in the even ($\ell=2,4,6,...$) zonal ($m=0$) harmonic coefficients $J_{\ell}$ of the multipolar expansion of the Newtonian part of the Earth's gravitational potential, which is not the most important source of aliasing precessions in the case of the Moon\cite{Wil03}, only $\delta J_2$ would be of some concern. Indeed, the mismodelled secular precessions induced by it on the lunar node and perigee amount to\footnote{The calibrated errors $\delta J_{\ell}$ of the EIGEN-CG01C Earth gravity field solution\cite{eigengrace01c} were used.} $-2.67\times 10^{-4}$ mas yr$^{-1}$ and $5.3\times 10^{-4}$ mas yr$^{-1}$, respectively; the impact of the other higher degree even zonals is negligible being $\leq 10^{-8}$ mas yr$^{-1}$.
As in the case of the spins, also the asphericity of the Moon has to be taken into account\cite{Bar75,Wex} according to
\eqI\dot\Omega_{J_2^{\rm Moon}}=-\rp{3}{2}\rp{n_{\rm Moon}\cos  F J^{\rm Moon}_2}{(1-e^2)^2}\left(\rp{R_{\rm Moon}}{a}\right)^2,\lb{precmoon}\eqF where $n_{\rm Moon}=\sqrt{GM(1+\mu)/a^3}$ is the lunar mean motion and $F$ is the angle between the orbital angular momentum and the Moon's spin angular momentum $\bm S_{\rm Moon}$; it is about 3.61 deg since the spin axis of the Moon is tilted by 1.54 deg  to the ecliptic and the orbital plane has an inclination of 5.15 deg to the ecliptic\cite{Ron05}.
Table \ref{tavolaMoon} and \rfr{precmoon} yield a mismodelled node precession due to $\delta J_2^{\rm Moon}$ of about 0.006  mas yr$^{-1}$, which is 6 times  larger than the Lense-Thirring rate.
%
%The bias due to $\delta J_2$ could be overcome by suitably combining the node and perigee according to
%\eqI \dot\Omega+k_1\dot\omega, \lb{combi}\eqF with
%\eqI k_1 =-\rp{\dot\Omega_{.2}}{\dot\omega_{.2}} =0.5030\eqF where $\dot\Omega_{.2}$ and $\dot\omega_{.2}$ are the coefficients of the secular %precessions of degree $\ell=2$ of the node and perigee, respectively. Their explicit expressions, along with those up to degree $\ell=20$, can be %found in \ct{Ior03}.
%The combination of \rfr{combi} is, by construction, insensitive by construction to the $\ell=2,\ m=0$ components of the static and time-varying %Earth's gravitational field.
%
For other sources of systematic errors induced by gravitational and even non-gravitational\cite{Vok99} perturbations see \ctz{Wil03} and references therein, especially \ctz{Cha83}.
%Note that the precessional effects considered there are referred to the ecliptic, not to the Earth's equator:
Among the N-body gravitational perturbations, the largest ones are due to the Sun's attraction.
In order to get an order-of-magnitude evaluation of their mismodelling, let us note that some of such effects are proportional to $n^2_{\oplus}/n_{\rm Moon}$; e.g. the node rate, referred to the equator, is\cite{Tap04}
\eqI \dot\Omega^{\rm \odot}=\rp{  3G\mathfrak{M}_{\odot}\cos I  }{  4a^3_{\oplus} n_{\rm Moon} }\left(\rp{3}{2}\sin^2\varepsilon-1\right)\approx -5\times 10^7\ {\rm mas\ yr}^{-1},\eqF where $\varepsilon=23.439$ deg is the obliquity of the ecliptic.
Since\cite{Sta98}  $\delta G\mathfrak{M}_{\odot}=5\times 10^{10}$ m$^3$ s$^{-2}$ and\cite{Groten} $\delta GM=8\times 10^5$ m$^3$ s$^{-2}$, we can assume a bias of $\approx 0.07$ mas yr$^{-1}$ which is 70 times larger than the Lense-Thirring precession.

Let us, now, consider the precision of LLR in reconstructing the lunar orbit with respect to the Lense-Thirring effect.
Concerning the precision in measuring the lunar precession rates, it amounts to about\cite{Mul91,Wil96,Mul07,Mul06} 0.1 mas yr$^{-1}$, i.e. it is two orders of magnitude larger than the Lense-Thirring precessions of \rfr{blu}.
The orbital perturbations experienced by a test particle are usually decomposed along three orthogonal directions of a frame co-moving with it; they are named radial $R$ (along the radius vector), transverse $T$ (orthogonal to the radius vector, in the osculating orbital plane) and normal $N$ (along the orbital angular momentum, out of the osculating orbital plane). According to, e.g., \ctz{Chr88}, the $R-T-N$ perturbations can be expressed in terms of the shifts in the Keplerian orbital elements as
\begin{equation}
\left\{
\begin{array}{lll}
\Delta R &=&\sqrt{ (\Delta a)^2 + \rp{[ (e\Delta a + a\Delta e)^2 + (ae\Delta{\mathcal{M}})^2 ]}{2}},\\\\
\Delta T &=& a\sqrt{ 1+\rp{e^2}{2} }\left[\Delta{\mathcal{M}}
+\Delta\omega+\cos I\Delta\Omega +\sqrt{(\Delta
e)^2+(e\Delta{\mathcal{M}})^2} \right],\\\\
\Delta N &=& a\sqrt{ \left(1+\rp{e^2}{2}\right)\left[\rp{(\Delta I
)^2}{2}+(\sin I\Delta\Omega)^2\right] },
%,
\end{array}
\right.
\end{equation}
where $\mathcal{M}$ is the mean anomaly. The lunar Lense-Thirring shifts
after one year are, thus
\begin{equation}
\left\{
\begin{array}{lll}
\Delta R_{\rm LT} &=& 0,\\\\
\Delta T_{\rm LT} &=& a\sqrt{ 1+\rp{e^2}{2}
}\left(\Delta\omega_{\rm LT}+\cos I\Delta\Omega_{\rm LT}\right)=- 0.38\ {\rm cm},\\\\
\Delta N_{\rm LT} &=& a\sqrt{1+\rp{e^2}{2}}\sin I\Delta\Omega_{\rm
LT}=0.07\ {\rm cm}.\lb{chris}
\end{array}
\right.
\end{equation}
It is important to  note that there is no Lense-Thirring secular signature in the Earth-Moon radial motion on which all of the efforts of LLR community have been concentrated so far. It can be shown that a short-period, i.e. not averaged over one orbital revolution, radial signal exists; it is proportional to\eqI \Delta r\propto \rp{2GS_{\oplus}}{c^2 n a^2}=2\times 10^{-5}\ {\rm m},\eqF which is too small to be detected since the present-day accuracy in estimating the amplitudes of radial harmonic signals is of the order of mm\cite{Mur07a}.
Major limitations come  from the post-fit Root-Mean-Square (RMS) accuracy with which the lunar orbit can be reconstructed; the present-day accuracy is about $2$ cm in the radial direction $R$ along the centers-of-mass of the Earth and the Moon\cite{Mul06}.
Improvements in the precision of the Earth-Moon ranging of the order of 1 mm are expected in the near future with the APOLLO program\cite{Wil04,Mur08}. Recently, sub-centimeter precision in determining range distances between a laser
on the Earth's surface and a retro-reflector on the Moon has been achieved\cite{Bat}.
However, it must be considered that the RMS accuracy in the $T$ and $N$ directions is likely worse than in $R$.
\section{Conclusions}
In this note we have examined the possibility of measuring the action of the intrinsic gravitomagnetic field of the spinning Earth on the lunar orbital motion with the LLR technique. After showing that the Lense-Thirring approximation is adequate for the Earth-Moon system, we found that the Lense-Thirring secular precessions of the Moon's node and the perigee induced by the Earth's spin angular momentum are of the order of $10^{-3}$ mas yr$^{-1}$ corresponding to transverse and normal secular shifts of $10^{-1}-10^{-2}$ cm yr$^{-1}$. The intrinsic gravitomagnetic field of the Earth does not secularly affect the radial component of the Moon's orbit;  a short-period, i.e. not averaged over one orbital revolution, radial oscillation is present, but its amplitude is of the order of $10^{-5}$ m. The current RMS accuracy in reconstructing the lunar orbit is of the order of cm in the radial direction; the harmonic components can be determined at the mm level. Forthcoming expected improvements in LLR should allow to reach the mm precision in the Earth-Moon ranging. The present-day accuracy in measuring the lunar precessional rate is of the order of $10^{-1}$ mas yr$^{-1}$. Major limitations come also from some orbital perturbations of classical origin like, e.g., the secular node precessions induced by the Sun and  the oblateness of the Moon which act as systematic errors and whose mismodelled parts are up to 70 times larger than the Lense-Thirring effects. As a consequence of our analysis, we are skeptical concerning the possibility of measuring intrinsic gravitomagnetism with LLR in a foreseeable future. The same conclusion holds also for the Lue-Starkman perigee precession predicted in the framework of the multidimensional braneworld DGP model of modified gravity; indeed, it is as large as the Lense-Thirring one for the Moon.

\section*{Acknowledgments}
I thank T. Murphy and J.G. Williams (NASA-JPL) for useful information and material supplied.

 %-----------------------------------------

\end{document}